\documentclass[prb,twocolumn,showkeys,showpacs,amsmath,amsfonts,floatfix,superscriptaddress,nofootinbib]{revtex4-1}

\usepackage{amssymb,amsmath,amstext}                
\usepackage{graphicx}                                               
\usepackage{epstopdf}                                               
\usepackage{color}       
\usepackage{bm}
\usepackage{appendix}                                              
\usepackage[latin2]{inputenc}
\usepackage{ulem}
\usepackage{latexsym}
\usepackage[colorlinks=true,citecolor=blue,linkcolor=magenta]{hyperref}
\def\be{\begin{equation}}
\def\ee{\end{equation}}
\def\bea{\begin{eqnarray}}
\def\eea{\end{eqnarray}}
\def\bi{\begin{itemize}}
\def\ei{\end{itemize}}

\newcommand{\bra}[1]{\mbox{$\langle #1 |$}}
\newcommand{\ket}[1]{\mbox{$| #1 \rangle$}}

\begin{document}

\title{ Simulation of many body localization and time crystals in two dimensions \\
                 with the neighborhood tensor update  }

\author{Jacek Dziarmaga}
\email{dziarmaga@th.if.uj.edu.pl}
\affiliation{Jagiellonian University, Institute of Theoretical Physics, 
             ul. \L{}ojasiewicza 11, 30-348 Krak\'ow, Poland }

\date{\today}

\begin{abstract}
The Heisenberg antiferromagnet with discrete disorder on an infinite square lattice is evolved in time from an initial N\'eel state. The simulation is performed with the neighborhood tensor update (NTU) algorithm for an infinite projected entangled pair state (iPEPS) [Phys. Rev. B 104, 094411 (2021)]. Ancillary spins are used to average over $2$ or $5$ discrete values of disorder. With a bond dimension up to $20$, evolution times are long enough to identify a many body localized regime for a strong enough disorder. Furthermore, the same Hamiltonian is subject to periodic spin flips. Simulations of the Floquet dynamics show that it can sustain a time crystalline stage for a strong enough disorder. 
\end{abstract}

\maketitle


\section{Introduction}
\label{sec:introduction}

Weakly entangled quantum states can be represented efficiently by tensor networks~\cite{Verstraete_review_08,Orus_review_14}, including the one-dimensional (1D) matrix product state (MPS)~\cite{fannes1992}, its two-dimensional (2D) generalization known as a projected entangled pair state (PEPS)~\cite{Nishino_2DvarTN_04,verstraete2004}, or a multi-scale entanglement renormalization ansatz~\cite{Vidal_MERA_07,Vidal_MERA_08,Evenbly_branchMERA_14,Evenbly_branchMERAarea_14}. The MPS ansatz provides a compact representation of ground states of 1D gapped local Hamiltonians~\cite{Verstraete_review_08,Hastings_GSarealaw_07,Schuch_MPSapprox_08} and purifications of their thermal states~\cite{Barthel_1DTMPSapprox_17}. It is also the ansatz underlying the density matrix renormalization group (DMRG)~\cite{White_DMRG_92, White_DMRG_93,Schollwock_review_05,Schollwock_review_11}. Analogously, the 2D PEPS is expected to represent ground states of 2D gapped local Hamiltonians~\cite{Verstraete_review_08,Orus_review_14} and their thermal states~\cite{Wolf_Tarealaw_08,Molnar_TPEPSapprox_15}, though representability of area-law states, in general, was shown to have its limitations~\cite{Eisert_TNapprox_16}. As a variational ansatz tensor networks do not suffer from the sign problem plaguing the quantum Monte Carlo. Consequently, they can deal with fermionic systems~\cite{Corboz_fMERA_10,Eisert_fMERA_09,Corboz_fMERA_09,Barthel_fTN_09,Gu_fTN_10}, as was shown for both finite~\cite{Cirac_fPEPS_10} and infinite PEPS (iPEPS)~\cite{Corboz_fiPEPS_10,Corboz_stripes_11}.

\begin{figure}[t!]
\vspace{-0cm}
\includegraphics[width=0.999\columnwidth,clip=true]{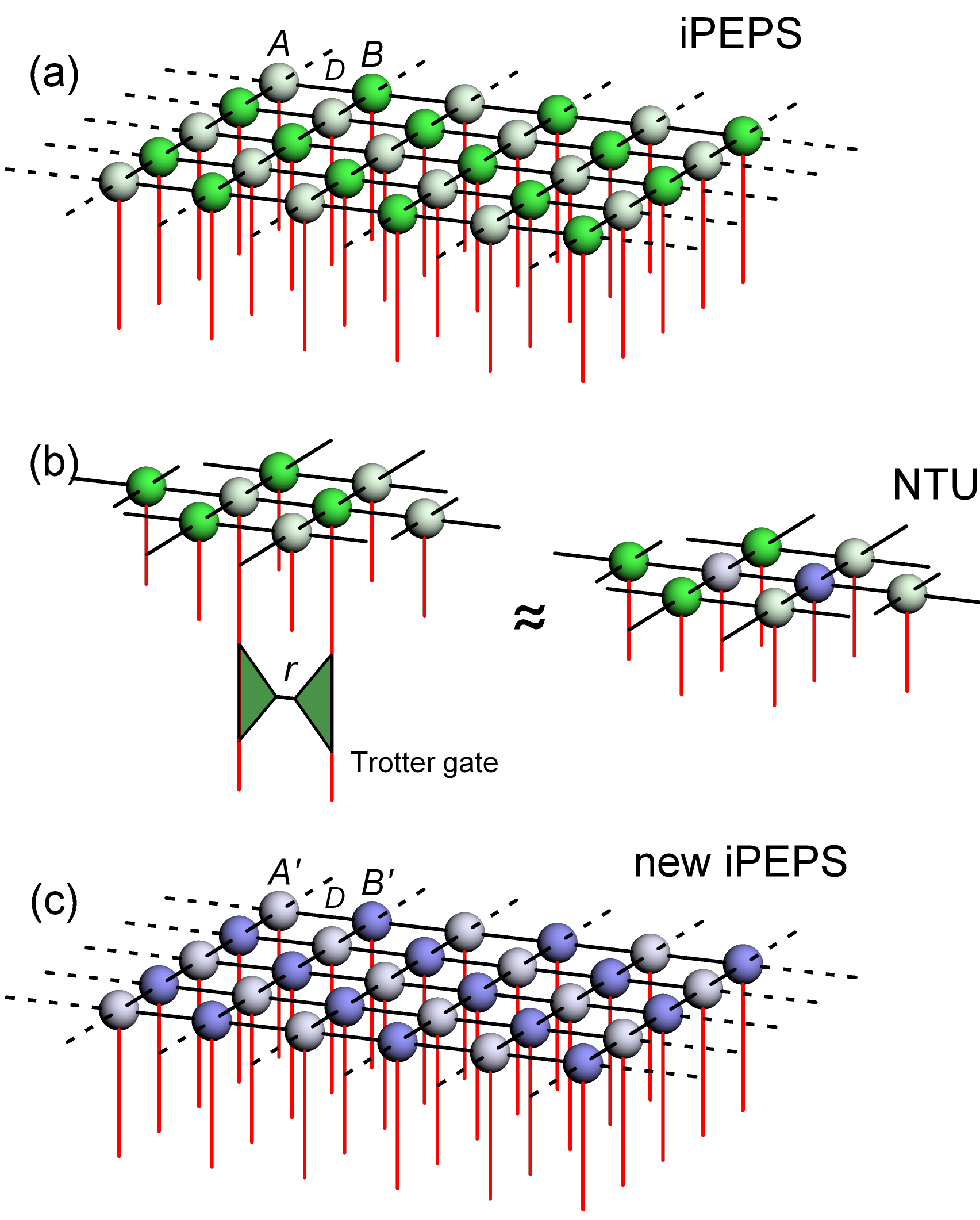}
\vspace{-0cm}
\caption{
{\bf Essential NTU. }
In (a) infinite PEPS with tensors $A$ (lighter green) and $B$ (darker green) on the two sublattices.
The red lines are physical spin indices and the black lines are bond indices, with bond dimension D, 
contracting NN sites. 
In one of Suzuki-Trotter steps a Trotter gate is applied to every horizontal NN pair of $A$-$B$ tensors
(but not to horizontal $B$-$A$ pairs). The gate can be represented as a contraction of two tensors by an index with dimension $r$. When the two tensors are absorbed into tensors $A$ and $B$ the bond dimension between them increases from $D$ to $r\times D$.
In (b) the $A$-$B$ pair -- with a Trotter gate applied to it -- is approximated by a pair of new tensors, $A'$ (lighter blue) and $B'$ (darker blue), connected by an index with the original dimension $D$. The new tensors are optimized to minimize the Frobenius norm of the difference between the two networks in (b).
The networks surround the considered NN bond with its six NN tensors. They provide a minimal tensor environment -- necessary to make efficient use of limited $D$ -- which can be contracted exactly in an efficient way.  
After $A'$ and $B'$ are converged, they replace all tensors $A$ and $B$ in a new iPEPS. 
Then the next Trotter gate can be applied.
More details of the NTU algorithm can be found in appendix \ref{app:algorithm}.
}
\label{fig:NTU}
\end{figure}

The PEPS was originally proposed as an ansatz for ground states of finite systems~\cite{Verstraete_PEPS_04, Murg_finitePEPS_07}, generalizing earlier attempts to construct trial wave-functions for specific models~\cite{Nishino_2DvarTN_04}. The subsequent development of efficient numerical methods for an infinite PEPS (iPEPS)~\cite{Cirac_iPEPS_08,Xiang_SU_08,Gu_TERG_08,Orus_CTM_09}, shown in Fig. \ref{fig:NTU}(a), promoted it as one of the methods of choice for strongly correlated systems in 2D. Its power was demonstrated, e.g., by a solution of the long-standing magnetization plateaus problem in the highly frustrated compound $\textrm{SrCu}_2(\textrm{BO}_3)_2$~\cite{matsuda13,corboz14_shastry}, establishing the striped nature of the ground state of the doped 2D Hubbard model~\cite{Simons_Hubb_17}, and new evidence supporting gapless spin liquid in the kagome Heisenberg antiferromagnet~\cite{Xinag_kagome_17}. Recent developments in iPEPS optimization~\cite{fu,Corboz_varopt_16,Vanderstraeten_varopt_16}, contraction~\cite{Fishman_FPCTM_17,Xie_PEPScontr_17}, energy extrapolations~\cite{Corboz_Eextrap_16}, and universality-class estimation~\cite{Corboz_FCLS_18,Rader_FCLS_18,Rams_xiD_18} pave the way towards even more complicated problems, including simulation of thermal states~\cite{Czarnik_evproj_12,Czarnik_fevproj_14,Czarnik_SCevproj_15, Czarnik_compass_16,Czarnik_VTNR_15,Czarnik_fVTNR_16,Czarnik_eg_17,Dai_fidelity_17,CzarnikDziarmagaCorboz,czarnik19b,Orus_SUfiniteT_18,CzarnikKH,wietek19,jimenez20,poilblanc20,CzarnikSS,Poilblanc_thermal}, mixed states of open systems~\cite{Kshetrimayum_diss_17,CzarnikDziarmagaCorboz,SzymanskaPEPS}, excited states~\cite{Vanderstraeten_tangentPEPS_15,ExcitationCorboz}, or real-time evolution~\cite{CzarnikDziarmagaCorboz,HubigCirac,tJholeHubig,Abendschein08,SUlocalization,SUtimecrystal,ntu,KZ2D,BH2Dcorrelationspreading}. 
In parallel with iPEPS, there is continuous progress in simulating systems on cylinders of finite width using DMRG. This numerically highly stable method that is now routinely used to investigate 2D ground states~\cite{Simons_Hubb_17,CincioVidal} was applied also to thermal states on a cylinder~\cite{Stoudenmire_2DMETTS_17,Weichselbaum_Tdec_18,WeichselbaumTriangular,WeichselbaumBenchmark,chen20}. However, the exponential growth of the bond dimension limits the cylinder's width to a few lattice sites. Among alternative approaches are direct contraction and renormalization of a 3D tensor network representing a 2D thermal density matrix \cite{Li_LTRG_11,Xie_HOSRG_12,Ran_ODTNS_12,Ran_NCD_13,Ran_THAFstar_18,Su_THAFoctakagome_17,Su_THAFkagome_17,Ran_Tembedding_18} and a cluster expansion \cite{cluster_thermal}.

In this paper we apply the recently defined neighborhood tensor update (NTU) algorithm \cite{ntu,KZ2D} to simulate unitary time evolution of a 2D many body localizing (MBL) system: the antiferromagnetic spin-$1/2$ Heisenberg model with discrete disorder initialized in the N\'eel state. The same simulatons were performed previously \cite{HubigCirac,SUlocalization} with the full update (FU) and simple update (SU) algorithms. NTU was intended as a reasonable trade off between FU and SU which is more accurate than SU but, at the same time, more efficient and stable than FU. The efficiency allows to reach bond dimensions up to $20$ that in turn permit longer evolution times. The longer evolution enables a more reliable identification of a MBL regime for strong enough disorder. Encouraged by this quantitative progress, we supplement the model with periodic spin flips. This Floquet dynamics was previously simulated with SU \cite{SUtimecrystal}. Here we obtain long enough evolution times, converged in the bond dimension, to identify time crystals for strong enough disorder.

This paper is organized as follows. In Sec. \ref{sec:ntu} we overview the NTU algorithm whose more technical details can be found in App. \ref{app:algorithm}. The applications follow in Sec. \ref{sec:mbl} and  \ref{sec:tc} to, respectively, many body localization and time crystals. We conclude in Sec. \ref{sec:conclusion}.


\section{Neighborhood tensor update}
\label{sec:ntu}

In the following NTU simulations we use the second order Suzuki-Trotter decomposition of small time steps. An application of a two-site NN Trotter gate is explained in a diagrammatic form in Fig. \ref{fig:NTU}. NTU is intermediate between the two most popular simulation schemes: the simple update (SU) and full update (FU) \cite{CzarnikDziarmagaCorboz}. In both after a Trotter gate is applied to a pair of nearest neighbor (NN) sites a bond dimension of the index between the sites is increased by a factor equal to the SVD rank of the gate. In order to prevent its exponential growth with time the dimension is truncated to a predefined value, $D$, in a way that minimizes error incurred by the truncation. The two schemes differ by a measure of the error: FU takes into account full infinite tensor environment while SU only the bonds adjacent to the NN sites. The former is expected to perform better in case of long range correlations while the latter is, at least formally, more efficient thanks to its locality though it makes less efficient use of the limited bond dimension than FU. In NTU the error measure is induced by the sites that are NN to the Trotter gate. It was demonstrated \cite{ntu} that in practice it compromises the FU accuracy only a little for a price of small numerical overhead over SU, hence it may turn out to be an optimal trade off for many applications, especially when quantum correlations are not exceedingly long like in e.g. Kibble-Zurek quenches in 2D \cite{KZ2D}. In this sense it seems to be tailored for time evolution of MBL systems where the localization is expected to limit the growth of correlations.  

NTU can be placed in a broader context by noting that it is a special case of a cluster update \cite{wang2011cluster} where the size of the tensor environment is a variable parameter interpolating between a local environment in SU and an infinite one in FU. In case of ground state calculations the cluster update was thoroughly investigated in Refs. \onlinecite{Lubasch_cluster_1,Lubasch_cluster_2} where an interplay between maximal achievable correlation length and the cluster size was demonstrated. In NTU the cluster includes the neighboring sites only, see Fig. \ref{fig:NTU}, to allow the error measure to be calculated exactly but with little numerical overhead over SU. The calculation involves only tensor contractions that are fully parallelizable. Its exactness warrants the error measure to be a manifestly Hermitian and non-negative quadratic form. This property is essential for stability of NTU and makes it distinct from FU where an approximate corner transfer matrix renormalization \cite{Orus_CTM_09,Orus_review_14} often breaks the Hermiticity and non-negativeness. In case of longer correlations the small environment can, admittedly,  make NTU converge with the bond dimension more slowly than FU but this can be often compensated by its better numerical efficiency and stability that allow NTU to reach higher bond dimensions \cite{ntu}.

\begin{figure*}[t!]
\vspace{-0cm}
\includegraphics[width=0.999\columnwidth,clip=true]{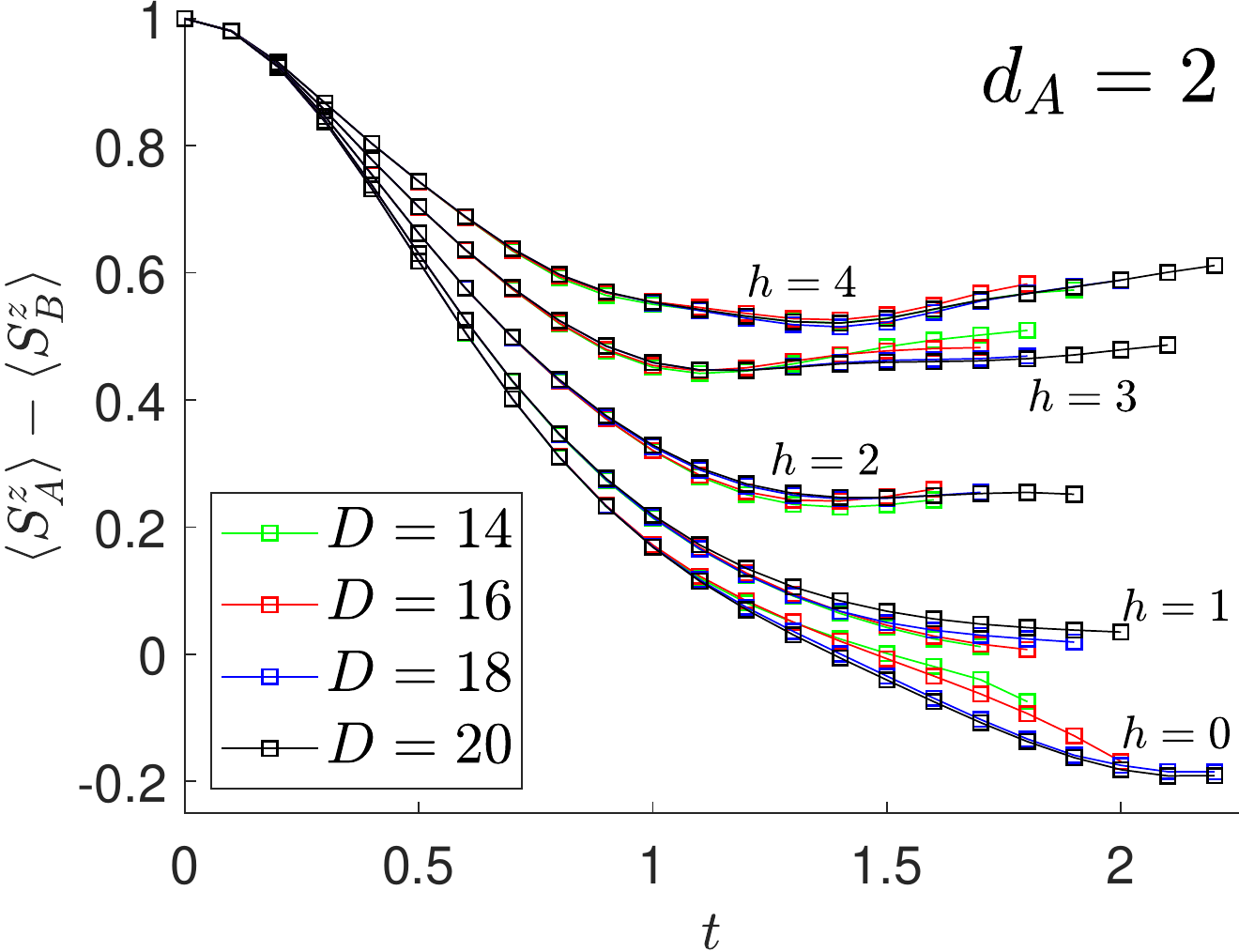}
\includegraphics[width=0.999\columnwidth,clip=true]{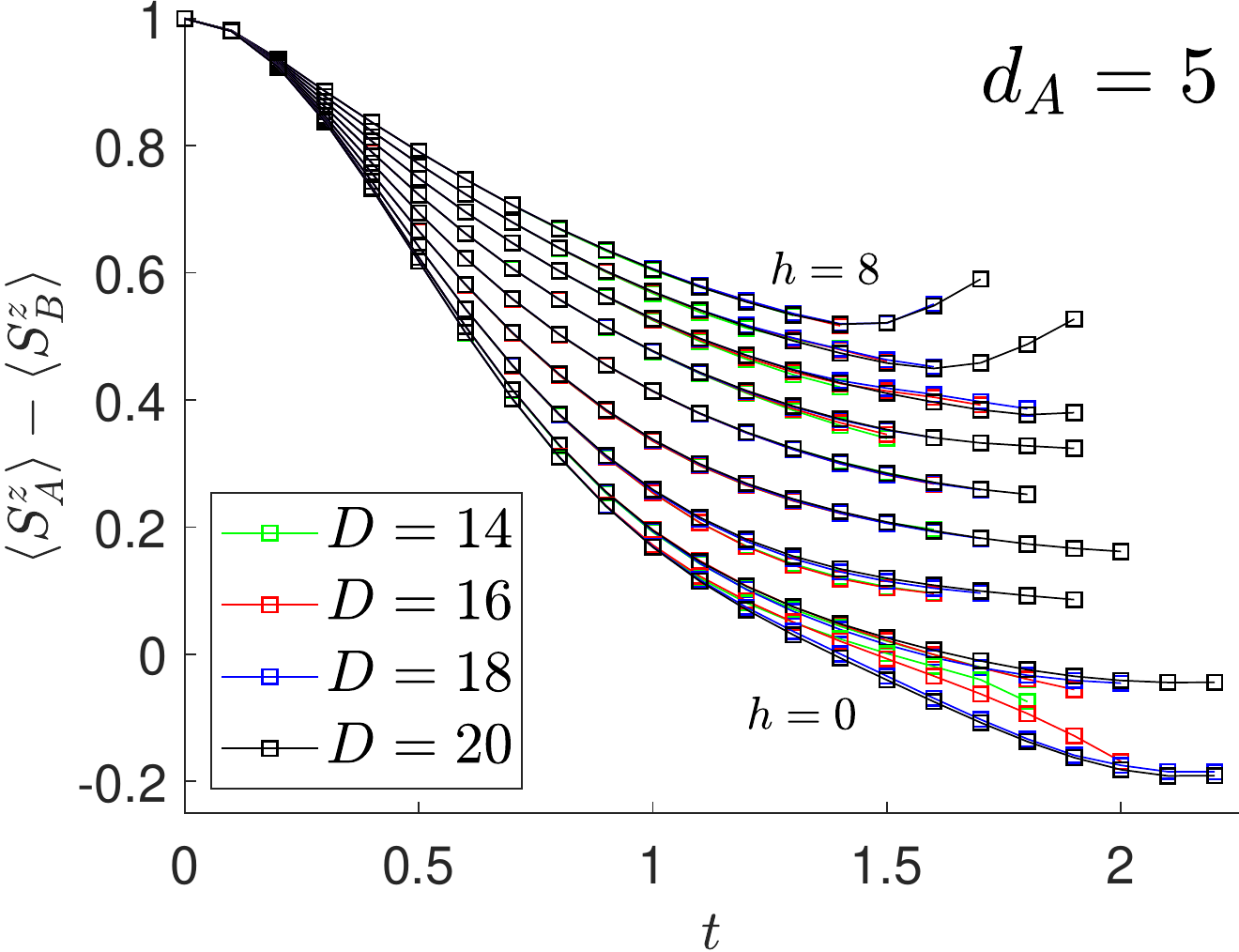}
\vspace{-0cm}
\caption{
{\bf Many body localization. }
A spin imbalance in function of time for different disorder strengths $h$ and ancillary dimensions $d_A$. 
Here the many body localizing Hamiltonian \eqref{HmblA} was initialized with the N\'eel state of spins \eqref{psi0AF}.
For $d_A=2$ and $5$ we show $h=0,1,..,4$ and $h=0,1,...,8$, respectively. 
The time step $dt=0.01$ and the evolution is terminated when the Trotter gate error \eqref{delta} exceeds $\delta=0.03$ for the first time. 
}
\label{fig:MBL}
\end{figure*}

\section{Many body localization}
\label{sec:mbl}

When excited the MBL systems evade thermalization but retaining memory of their initial conditions \cite{MBL_Ann_Phys,RMP_MBL_thermalization}. The memory is sustained by localization that gives rise to quasi-local constants of motion \cite{MBL_Pal_Huse,MBL_Ronen_Huse_Altman,MBL_Gluza_Eisert} that allow the evolving state to equilibrate but to a non-thermal stationary state. In 2D which is harder to investigate by any means --- analytic, numerical or experimental \cite{MBL_2D_exp} --- the issue is far from settled. The 2D MBL was suggested to be unstable towards a crossover to ergodicity on very long time scales \cite{MBL_2D_question}. Several numerical approaches were proposed to address the 2D problem \cite{MBL_2DBH,MBL_Simon,peps2018MBL,HubigCirac,Pollmann_MBL,disorder_ancilla_1,disorder_ancilla_2,SUlocalization}. This paper follows the approach of simulating the time evolution of a 2D MBL system with the iPEPS tensor network \cite{peps2018MBL,HubigCirac,disorder_ancilla_1,disorder_ancilla_2,SUlocalization} but it employs the NTU algorithm \cite{ntu}. For a localizing system this algorithm is expected to offer an optimal combination of stability and efficiency.

In order to be more specific, we consider the antiferromagnetic Heisenberg model with disorder on an infinite square lattice:
\be 
H_{\rm } =
\sum_{\langle j,j'\rangle} \vec S_j \vec S_{j'} + \sum_j h_j S^z_j.
\label{Hmbl}
\ee
Here $\vec S_j=\frac12\vec\sigma_j$ is a spin-$1/2$ operator at site $j$. Every random $h_j$ is drawn independently from a set of $d_A$ evenly spaced discrete values in an interval $[-h,h]$. For instance, $h_j\in \{-h,h\}$ for $d_A=2$, $h_j\in \{-h,0,h\}$ for $d_A=3$, etc. The Hamiltonian is initialized with a N\'eel state, $\ket{\text{N\'eel}}$, with spins pointing up/down along the $z$-axis on sublattice A/B of an infinite checkerboard lattice. This initial product state is evolved in time. Expectation values of spin operators at time $t$ are averaged over all possible realisations of the disorder.

Here as in Refs. \onlinecite{HubigCirac,disorder_ancilla_1,disorder_ancilla_2,SUlocalization} an equivalent formulation of the problem is considered:
\be 
H_{\rm MBL} =
\sum_{\langle j,j'\rangle} \vec S_j \vec S_{j'} + \frac{h}{S} \sum_j A^z_{j} S^z_j,
\label{HmblA}
\ee
where $A^z_{j}$ is an ancilla spin-$S$ operator at site $j$ that takes $d_A=2S+1$ evenly spaced discrete values in an interval $[-S,S]$. In this formulation the initial state is replaced by an equivalent uniform product:
\be 
\ket{\psi(0)} = \ket{\text{N\'eel}} \prod_{j} \ket{+}_j,
\label{psi0AF}
\ee 
where $j$ runs over ancillas. Every ancillary spin-$S$ is prepared in a superposition:
\be 
\ket{+}_j=(2S+1)^{-1/2}\sum_{m_j=-S}^{S} \ket{m_j},
\ee 
with the same probability for all evenly spaced values of disorder $h_j=m_j h/S\in[-h,h]$. The two formulations are equivalent in the sense that a disorder-averaged expectation value of any operator $\cal{O}$ at time $t$ in the former is equal to $\bra{\psi(t)}\cal{O}\ket{\psi(t)}$ in the latter. 

The Hamiltonian \eqref{HmblA} is the same as in Ref. \onlinecite{SUlocalization} and the same ancillary dimensions $d_A=2$ and $5$ are considered for a range of disorder strengths. Unlike Ref. \onlinecite{SUlocalization} NTU is employed instead of SU and bond dimensions up to $D=20$ are reached instead of $D=4,5$. As a result NTU evolution times are long enough to reach some semi-quantitative conclusions without any extrapolation in time.

\begin{figure*}[t!]
\vspace{-0cm}
\includegraphics[width=0.999\columnwidth,clip=true]{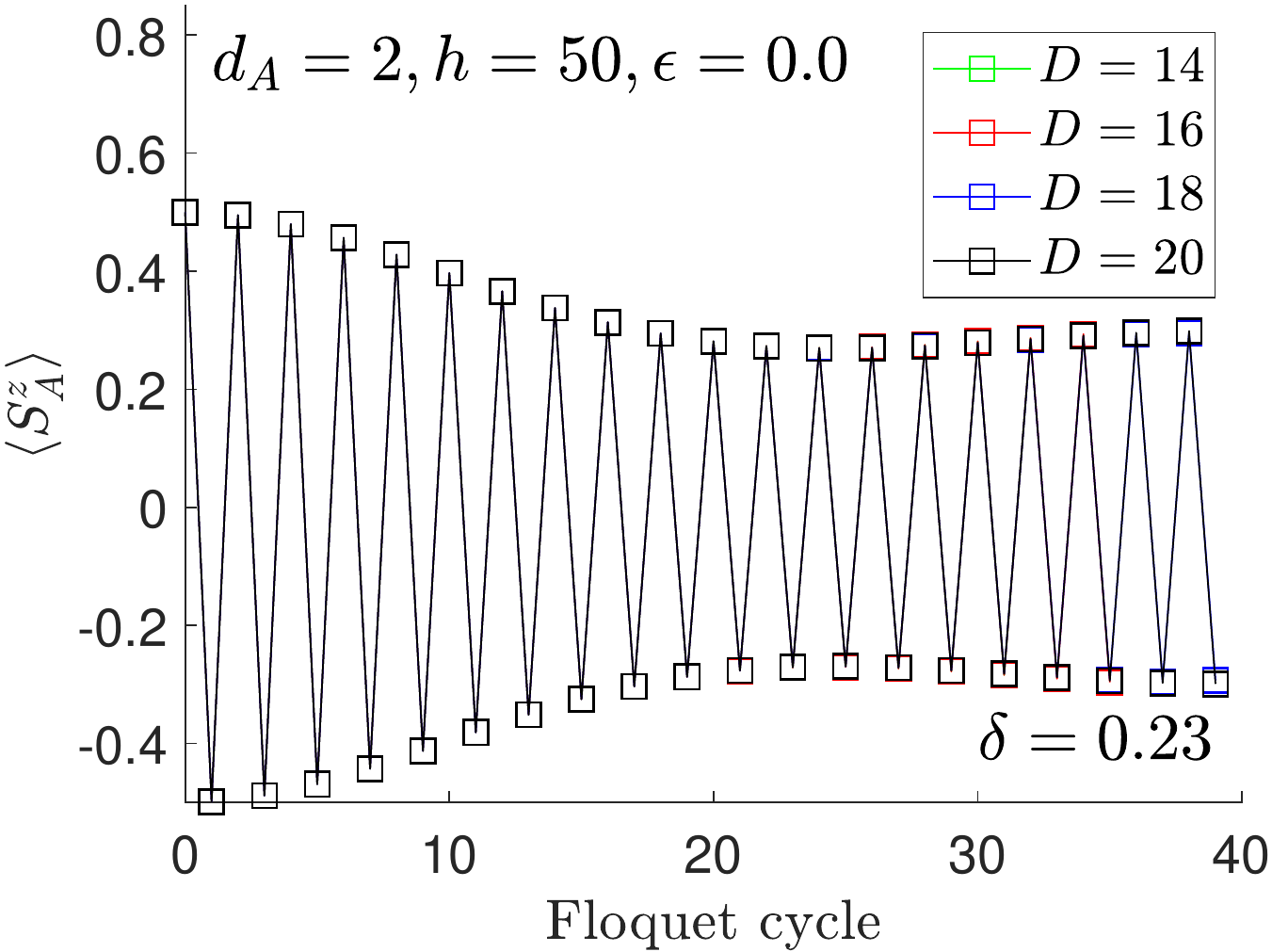}
\includegraphics[width=0.999\columnwidth,clip=true]{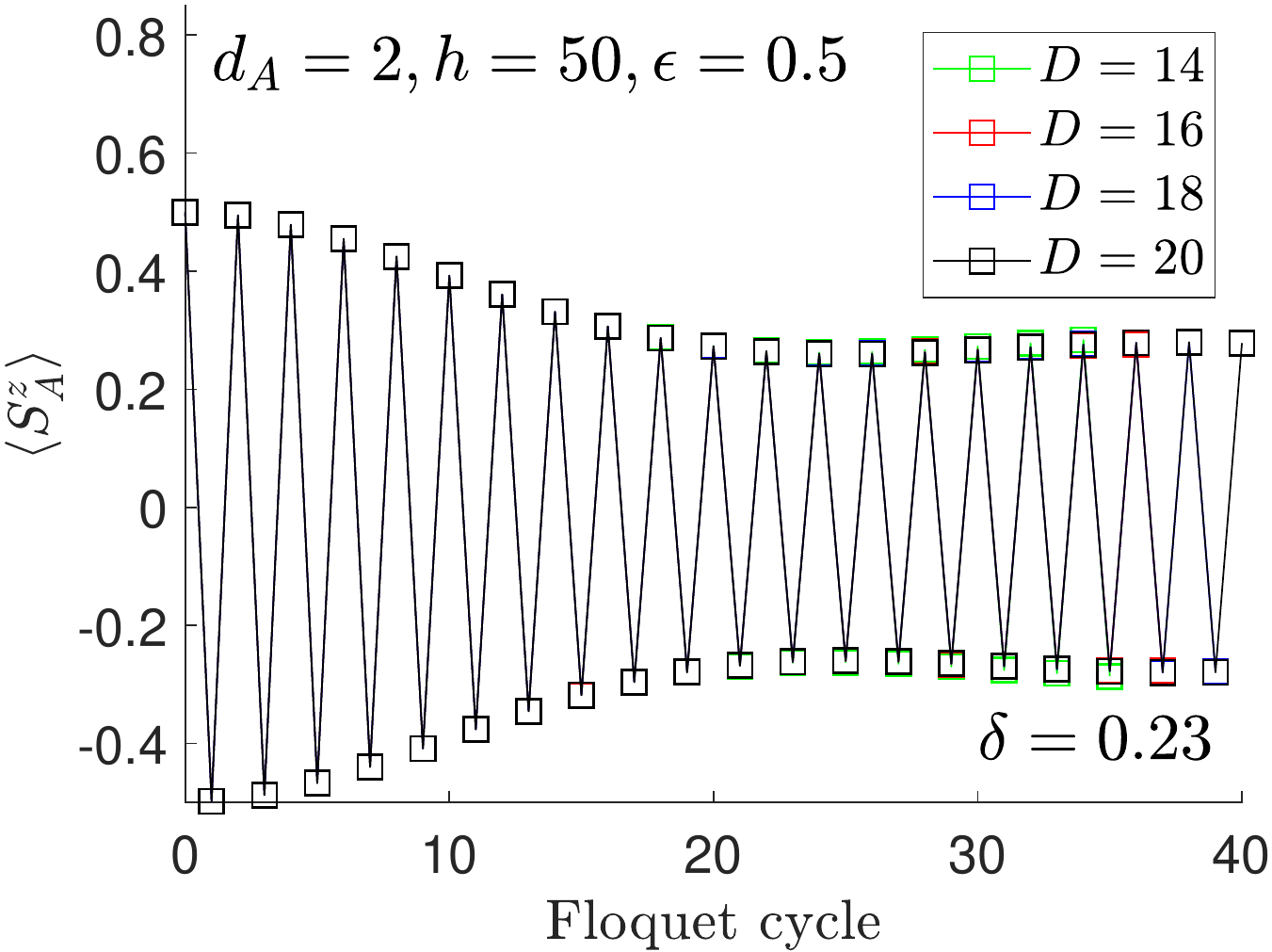}
\includegraphics[width=0.999\columnwidth,clip=true]{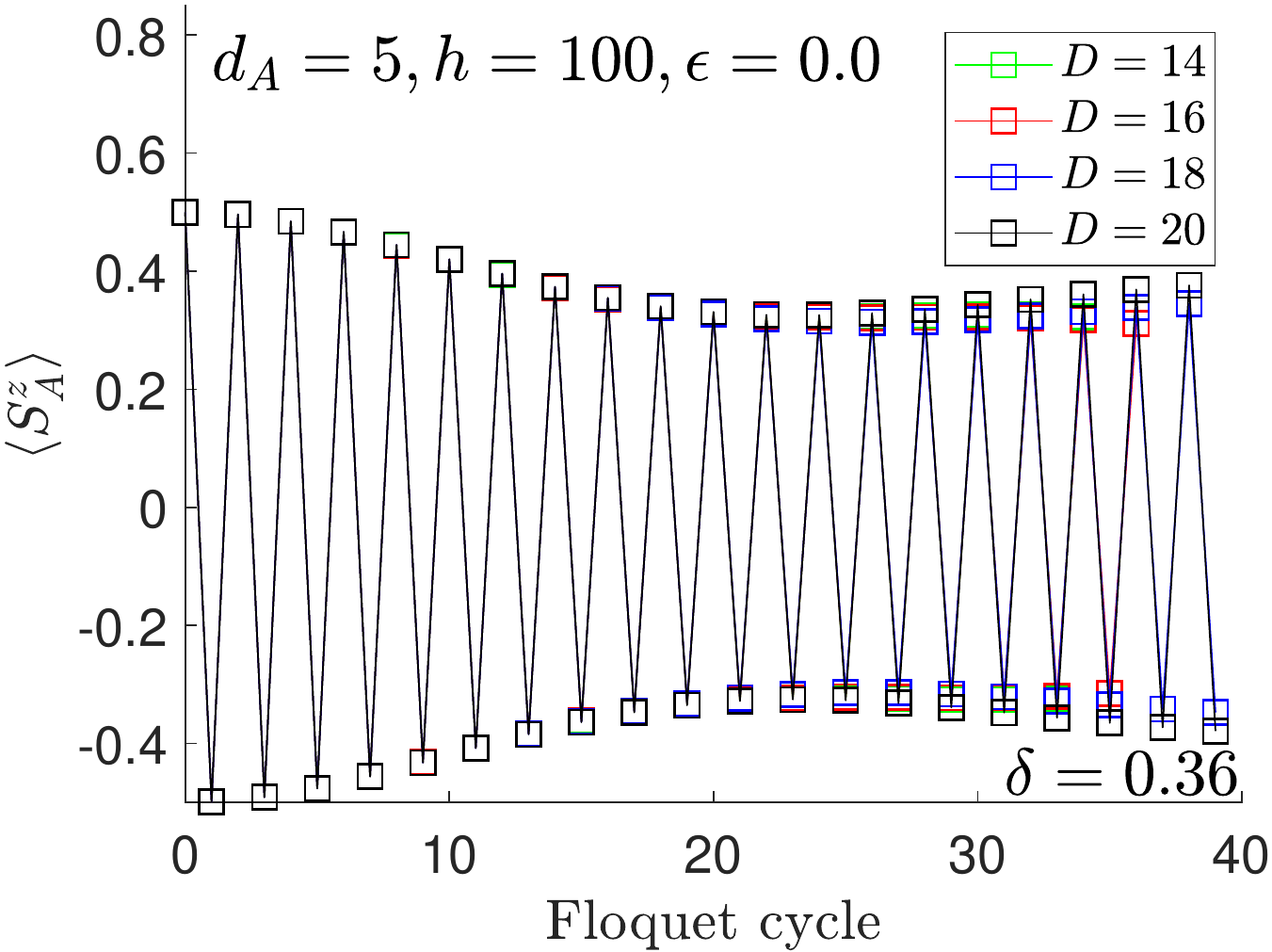}
\includegraphics[width=0.999\columnwidth,clip=true]{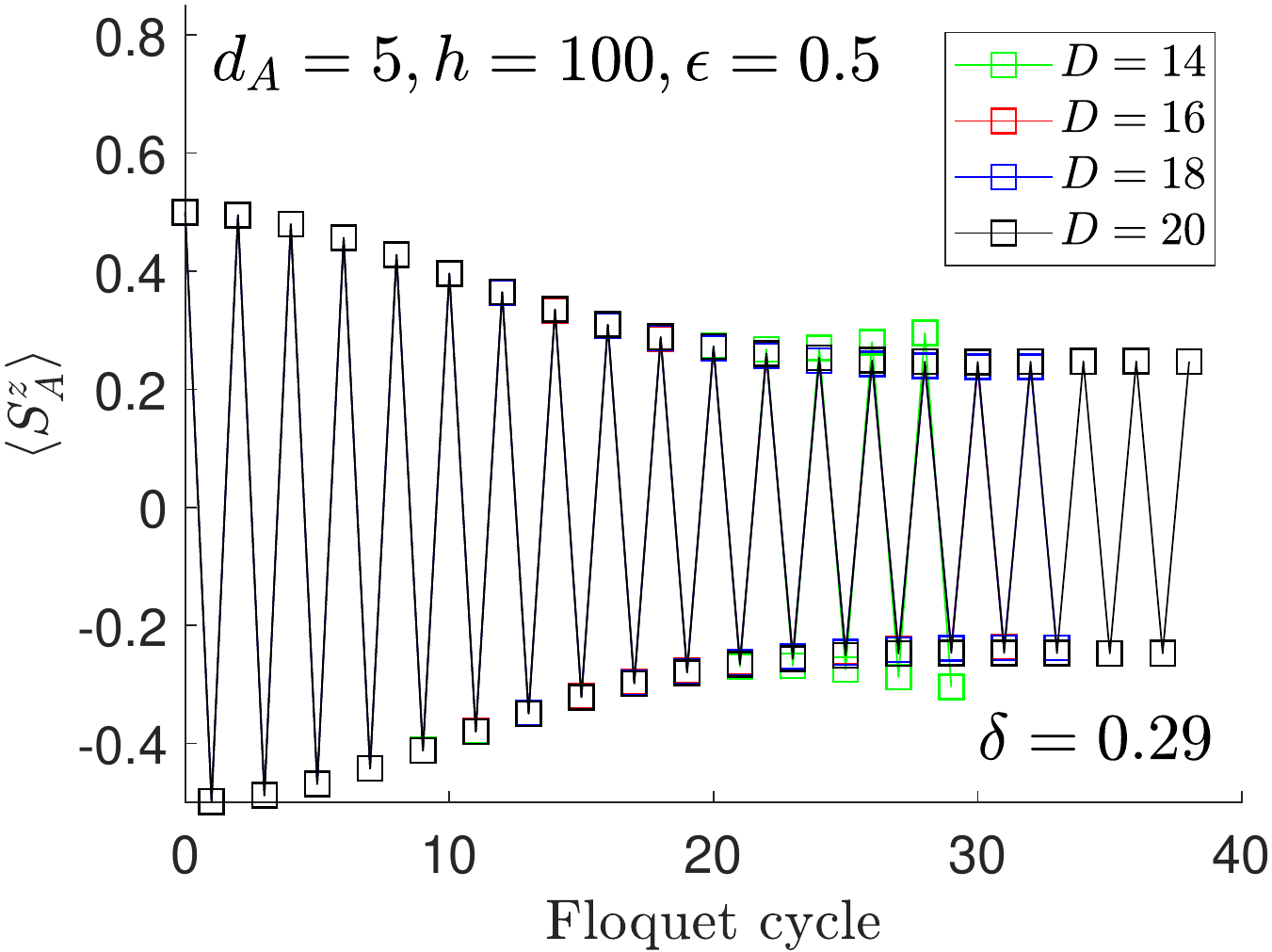}
\vspace{-0cm}
\caption{
{\bf Time crystallization. }
A stroboscopic time evolution of magnetization on sublattice $A$ in function of number of cycles for two combinations $(d_A,h)$ of ancillary dimension and disorder strength, $(2,50)$ and $(5,100)$, and two values of the angle deficit parameter: $\epsilon=0$ and $0.5$. 
Here the Floquet Hamiltonian \eqref{HF} with period $T=0.1$ was initialized with a N\'eel state of spins along $S^Z$ \eqref{psi0AF}.
The time step was $dt=0.001$ and the evolution was terminated when the Trotter gate error $\delta$, defined in \eqref{delta}, exceeded for the first time the value in the bottom-right corner of each panel. In the top and bottom-right panels this value is chosen such that the results for $D=18,20$ appear converged until the termination. In the bottom-left case there is still some dependence on $D$ with a tendency to improve the time crystal with increasing $D$.  
}
\label{fig:TC}
\end{figure*}

Figure \ref{fig:MBL} shows time evolution of spin imbalance between sublattices $A$ and $B$, 
$\bra{\psi(t)}S^z_A\ket{\psi(t)}-\bra{\psi(t)}S^z_B\ket{\psi(t)}$, in function of evolution time $t$ for several values of disorder strength $h$. 
In particular both panels show the evolution without disorder, $h=0$, when the imbalance crosses from positive to negative. Not quite surprisingly, in the absence of MBL this simulation becomes poorely converged in $D$ when the system begins to thermalize and growing entanglement becomes difficult to accommodate by the tensor network. The same comment, though to a lesser extent, applies to a finite but weak disorder with $h=1$. 
For higher disorder strengths the difference between $d_A=2$ and $d_A=5$ appears to be rather quantitative. In the former case, as far as the limited evolution time permits to conclude, there seems to be a crossover to MBL between $h=1$ and $h=2$. For $h\geq2$ the spin imbalance at first goes down from the initial $1$ before it bounces up. The memory of the initial spin imbalance between the sublattices is not lost. In the case of $d_A=5$ a similar bounce up is observed, within available evolution time, only for the highest disorder strengths $h=7,8$. For $h=2,...,6$ the NTU evolution terminates before bouncing up. It remains an open question whether the spin imbalance is bouncing up at later times or the evolution is heading straight towards thermalization and how precisely it depends on $h$. For the available evolution times we can crudely estimate the crossover between the (hypothetical) thermalization and MBL to take place between $h=1$ and $h=7$. 

It is worth emphasizing that, thanks to NTU (instead of SU) and bond dimensions up to $D=20$ (instead of $D=4,5$), our ($D$-converged) evolution times are typically twice as long as in Ref. \onlinecite{SUlocalization}. This allows us to identify the MBL regime not only for $d_A=5$ but also for the two-level disorder. 

In the next section the same Hamiltonian \eqref{HmblA} is promoted to a Floquet Hamiltonian by the action of time-periodic spin flips. 

\begin{figure*}[t!]
\vspace{-0cm}
\includegraphics[width=0.999\columnwidth,clip=true]{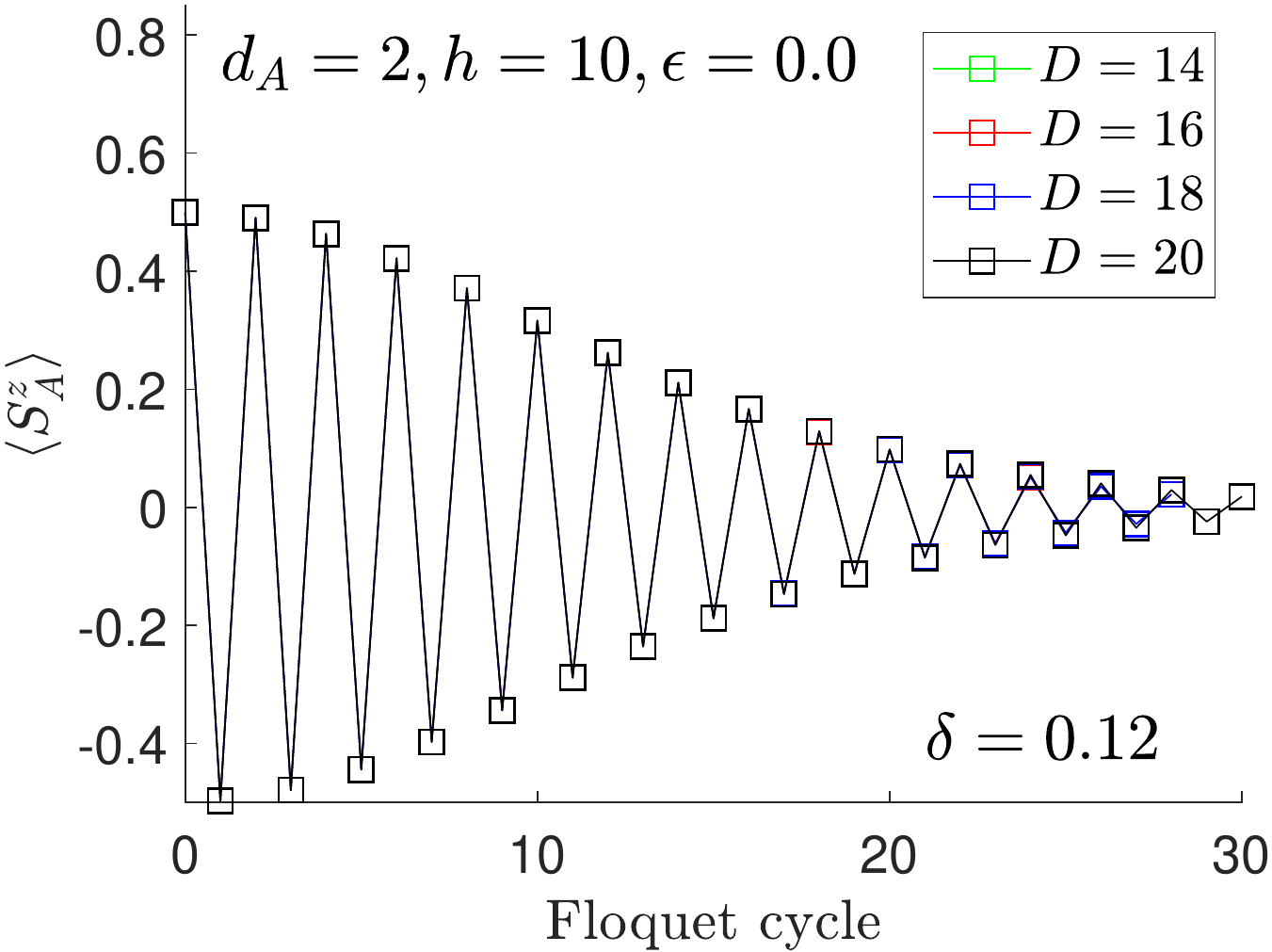}
\includegraphics[width=0.999\columnwidth,clip=true]{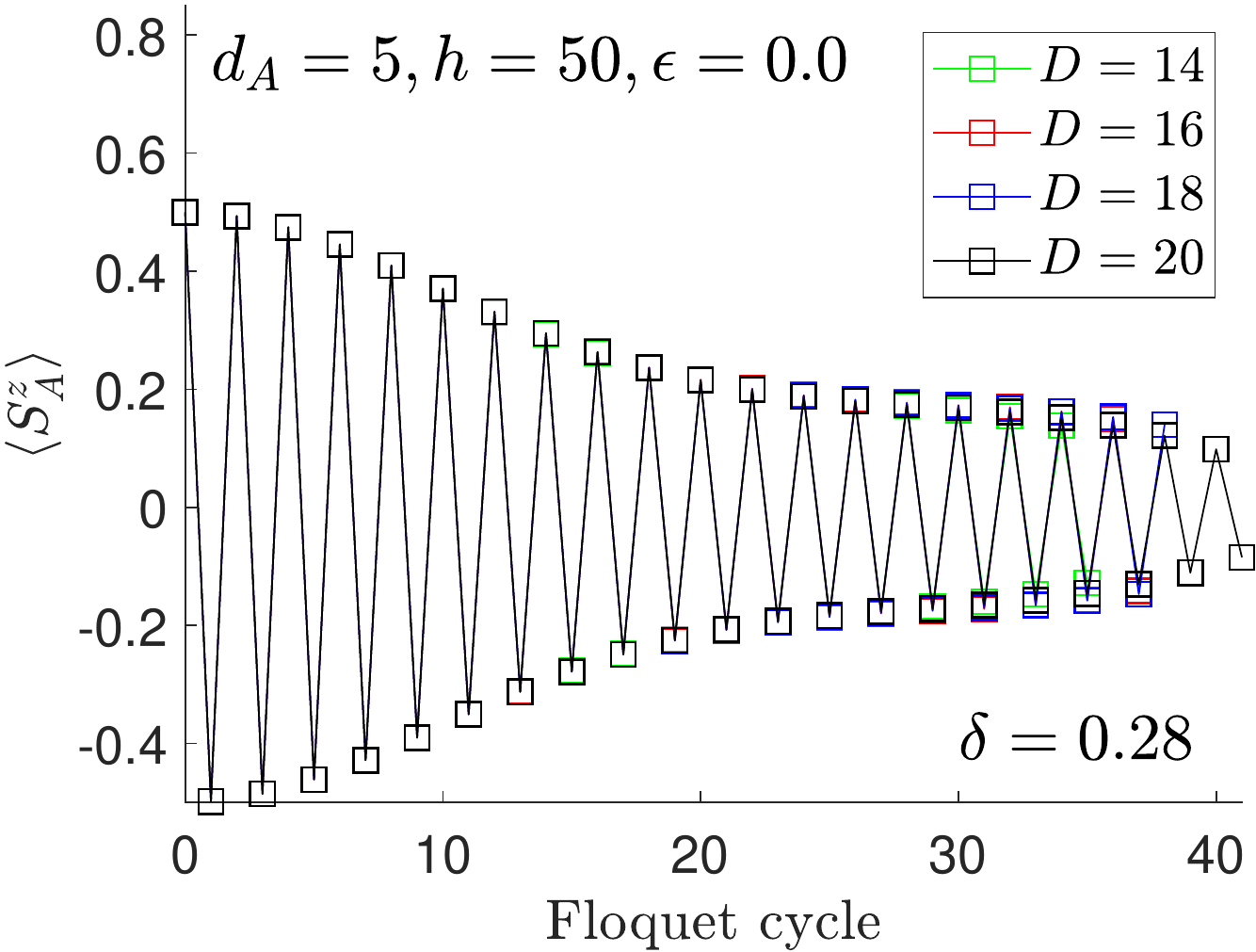}
\vspace{-0cm}
\caption{
{\bf Time crystalization failure. }
Two generic examples of stroboscopic time evolution where the disorder is too weak to sustain a time crystal (TC) that extends beyond the achievable simulation time. In the left panel TC cannot be identified at all while in the right one there seems to be a transient TC stage that begins to disappear after around $40$ Floquet cycles.
Here the Floquet Hamiltonian \eqref{HF} with period $T=0.1$ was initialized with the N\'eel state of spins \eqref{psi0AF}.
The time step was $dt=0.001$ and the evolution was terminated when the Trotter gate error $\delta$, defined in \eqref{delta}, exceeded for the first time the value in the bottom-right corner of each panel. 
%
}
\label{fig:TCbis}
\end{figure*}

\section{Time crystals}
\label{sec:tc}

Time crystals were at first envisioned by Wilczek as spontaneous time translation symmetry breaking in a ground state of a quantum Hamiltonian \cite{TC_Wilczek} but this original idea was proved wrong by a no go theorem \cite{TC_Watanabe_Oshikawa,TC_Kozin_Kyriienko}. More recently spontaneous discrete time translation symmetry breaking was demonstrated \cite{TC_Sacha_2015,TC_Khemani,TC_Nayak,TC_Yao} in Floquet systems that are periodic in time. These ideas inspired experiments that were performed in a variety of physical platforms \cite{Zhang2017,Choi2017,Barrett2018,Sreejith2018,Barrett2018PRB,vanderStraten2018,Stoof2019,Smits_2020}. The fast growing field is already subject to several reviews \cite{Sacha_review,Khemani_review,Yao_review}. A few mechanisms were proposed how a Floquet system can avoid heating. They include the many body localization \cite{Ponte2015,Lazarides2015,Abanin2016}, prethermalization \cite{Kuwahara2016,Canovi2016,Machado2019}, and many body quantum scars \cite{Turner2018,Michailidis2020,Pizzi2020,Surace2021}. 

Here as in Ref. \onlinecite{SUtimecrystal} the MBL mechanism is considered. The Floquet Hamiltonian with period $T$ is 
\be 
H_{\rm{F}}=
\left\{  
\begin{array}{ll}
H_{\rm MBL},  &, \rm{when~} t\in[0,T/2), \\
H_{\rm f},    &, \rm{when~} t\in[T/2,T).
\end{array}
\right.
\label{HF}
\ee 
Here $H_{\rm MBL}$ is the same as in \eqref{HmblA} and 
\be 
H_f = (2\pi/T-2\epsilon) \sum_j S^x_j.
\ee 
For $\epsilon=0$ its evolution operator, $e^{-i H_{\rm f}T/2}$, is a perfect spin flip operator that is applied once per every period $T$. In general it is an imperfect spin rotation with an angle deficit $\epsilon T$. An initial state is again the N\'eel state in \eqref{psi0AF}. 

Figure \ref{fig:TC} shows examples of stroboscopic time evolutions for ancillary dimensions $d_A=2,5$ and corresponding disorder strengths $h=50,100$ that turn out to be strong enough to stabilize an unambiguous time crystalline stage. The same figure demonstrates robustness of these time crystals against the spin-flip imperfection with $\epsilon=0.5$. On the other hand, when the disorder is not strong enough then the time crystalline stage either cannot be identified at all or disappears within the achievable evolution time, see the left and right panels in Fig. \ref{fig:TCbis}, respectively. The disappearance is accompanied by a worsening convergence with $D$ as might have been expected when MBL is not effective enough. 

\section{Conclusion}
\label{sec:conclusion}

The neighborhood tensor update algorithm \cite{ntu} was employed to simulate unitary time evolution of the square-lattice antiferromagnetic spin-$1/2$ Heisenberg model with discrete disorder. Starting from the N\'eel state many body localized regimes were identified --- for both the 2-level and the 5-level disorder --- where the system retains some memory of the initial state. When promoted to a Floquet Hamiltonian with periodic (imperfect) spin flips the simulations revealed extended time-crystalline stages for strong enough disorder. 

\acknowledgements
%
I am indebted for stimulating comments from Marin Bukov, Piotr Czarnik, Marek Rams, and Aritra Sinha.
This research was supported in part by the National Science Centre (NCN), Poland under project 2019/35/B/ST3/01028.
%
\appendix

\begin{figure}[t!]
\vspace{-0cm}
\includegraphics[width=0.9999\columnwidth,clip=true]{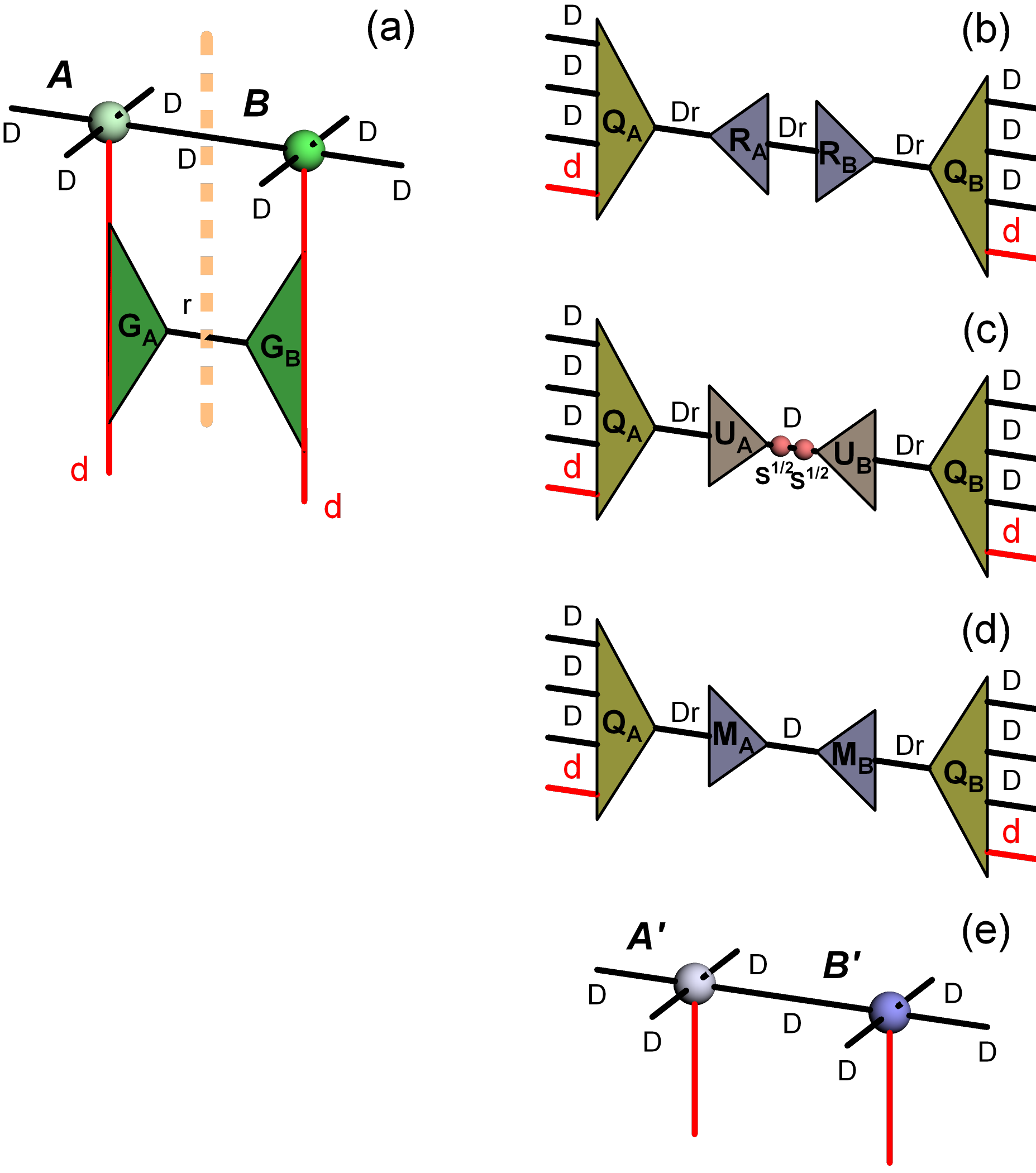}
\vspace{-0cm}
\caption{
{\bf Trotter gate. }
In (a) 
a 2-site gate is applied to physical indices of NN tensors $A$ and $B$ as in Fig. \ref{fig:NTU} (b). The gate is replaced by two tensors, $G_A$ and $G_B$, contracted by an index with dimension $r$.
In (b)
the tensor contraction $A\cdot G_A$ is QR-decomposed into $Q_AR_A$. Similarly $B\cdot G_B=Q_BR_B$. 
Isometries $Q_{A,B}$ will remain fixed.
In (c)
after SVD, $R_AR_B^T=U_ASU_B^T$, $S$ is truncated to $D$ leading singular values.
In (d)
matrices $M_A=U_AS^{1/2}$ and $M_B^T=S^{1/2}U_B^T$ are made by absorbing a square root of truncated $S$ symmetrically.
In (e)
at this point new iPEPS tensors could be obtained as $A'=Q_A\cdot M_A$ and $B'=Q_B\cdot M_B$ ending the story.
This scheme was referred to as an SVD update (SVDU) in Ref. \onlinecite{ntu}. 
In NTU scheme matrices $M_{A,B}$ are further optimized in the neighborhood tensor environment in Fig. \ref{fig:NTU} before being contracted back with the fixed isometries $Q_{A,B}$ to make new iPEPS tensors $A'$ and $B'$.
}
\label{fig:2site2}
\end{figure}

\begin{figure}[t!]
\vspace{-0cm}
\includegraphics[width=0.80\columnwidth,clip=true]{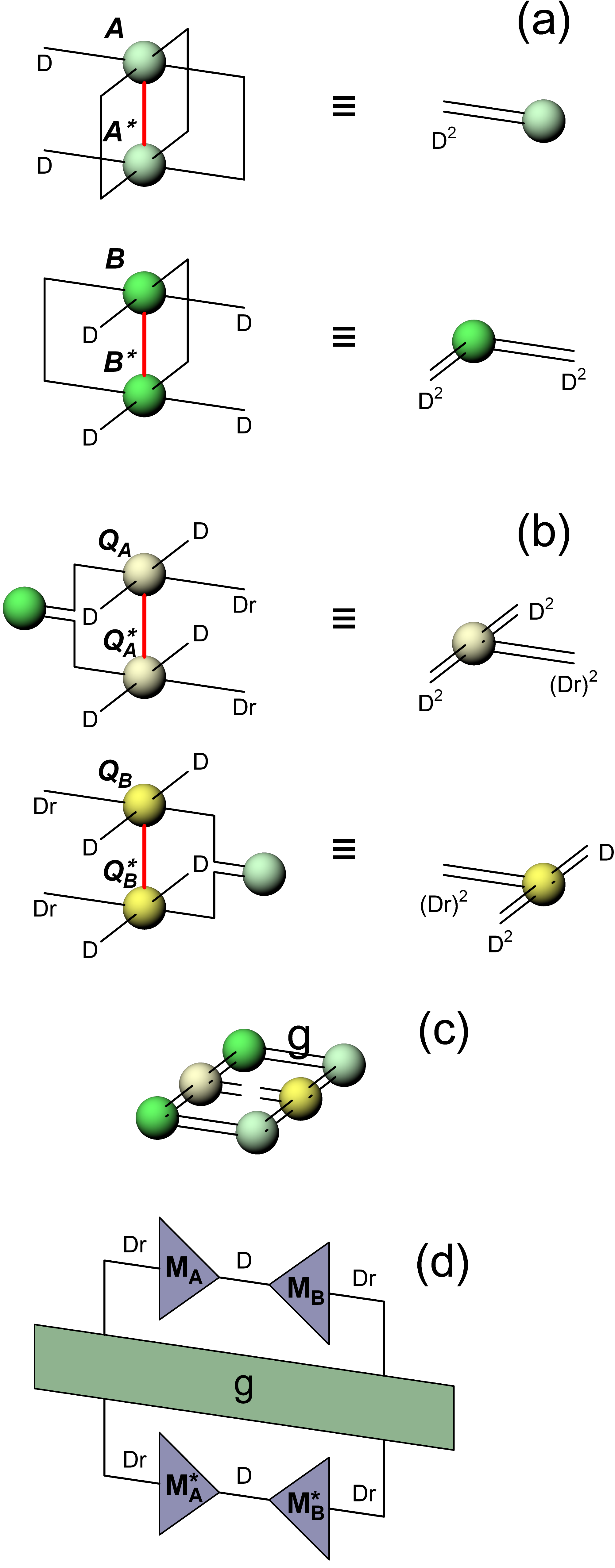}
\vspace{-0cm}
\caption{
{\bf Metric tensor. }
Norm squared of the matrix product, $\vert\vert M_AM_B^T \vert\vert^2$, is shown in (d).
Here $g$ is a metric tensor assembled in (c). 
The upper/lower pair of free indices in (c) corresponds to the upper/lower pair of indices of $g$ in (d).
Diagram (c) is assembled from 6 edge tensors -- two examples of which are shown in (a) -- 
and 2 double isometric tensors in (b).
Cost of all the contractions scales like $D^8$ and is fully parallelizable.  
}
\label{fig:NTUenv}
\end{figure}

\section{Algorithm}
\label{app:algorithm}

\begin{figure}[h]
\vspace{-0cm}
\includegraphics[width=0.5\columnwidth,clip=true]{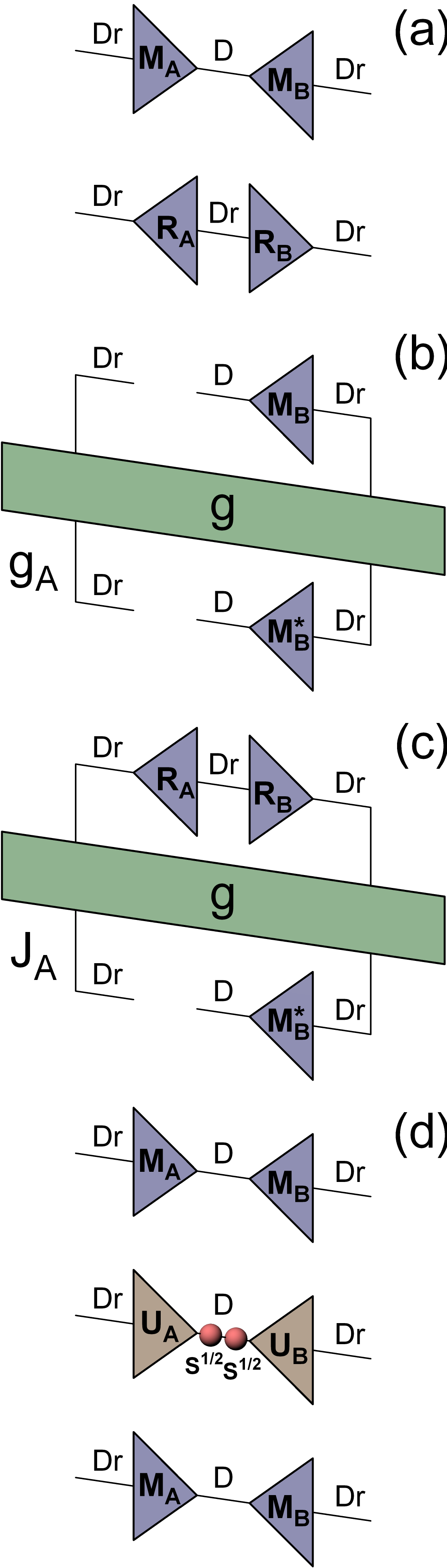}
\vspace{-0cm}
\caption{
{\bf NTU optimization loop. }
In (a)
matrices $M_A, M_B$ are optimized for their product, $M_A M_B^T$, to be the best approximation to the exact product, $R_AR_B^T$. The error is measured with the metric in Fig. \ref{fig:NTUenv}(d). 
In (b)
reduced metric tensor $g_A$ for matrix $M_A$.
In (c)
reduced source term $J_A$ for matrix $M_A$.
In (d) 
a product of converged matrices is subject to a SVD, $M_A M_B=U_ASU_B^T$, after which new balanced matrices, $M_A=U_AS^{1/2}$ and $M_B^T=S^{1/2}U_B^T$, are formed by absorbing singular values $S$ in a symmetric way. 
However, iterative optimization of the matrices is not symmetric. Before optimization with respect to $M_A$ the matrices are ``tilted'' as $M_A=U_AS$ and $M_B^T=U_B^T$ and vice versa \cite{Evenbly2018}. 
}
\label{fig:NTUg}
\end{figure}

The core of NTU scheme \cite{ntu} is outlined in figures \ref{fig:NTU}, \ref{fig:2site2}, \ref{fig:NTUenv}, and \ref{fig:NTUg}. 
Figure \ref{fig:2site2} shows application of a 2-site Trotter gate to a pair of NN iPEPS tensors, $A$ and $B$. The rank-$r$ gate enlarges the bond dimension from $D$ to $Dr$ that has to be truncated back to $D$ in an optimal way. For the sake of efficiency, a QR decomposition is introducing reduced matrices $R_A$ and $R_B$ in place of full tensors \cite{Evenbly2018}. A SVD of their product, $R_AR_B^T$, can be truncated to provide a preliminary truncation of the bond dimension. The truncated reduced matrices, $M_A$ and $M_B$, are further optimized iteratively in Figs. \ref{fig:NTUenv} and \ref{fig:NTUg}. After convergence they are contracted again with the fixed isometries, $Q_A$ and $Q_B$, into new iPEPS tensors $A'$ and $B'$. The numerical cost of the procedures in Fig. \ref{fig:2site2} scales as $D^5$.

The further NTU optimization minimizes the difference between the LHS and the RHS of the equation in Fig. \ref{fig:NTU}(b). The contraction $A'-B'$ on its RHS (the purple tensors) is the same contraction as in Fig. \ref{fig:2site2}(e) and it has to be understood as the diagram in Fig. \ref{fig:2site2}(d). This way the RHS depends on the product of the matrices to be optimized: $M_AM_B^T$. Therefore, the norm squared of the difference between the LHS and the RHS can be written as 
\be 
F\left(M_AM_B^T\right)=
\left[M_AM_B^T-R_AR_B^T\right]^\dag ~ g ~ \left[M_AM_B^T-R_AR_B^T\right],
\label{cost}
\ee 
where $g$ is the metric tensor constructed in Fig. \ref{fig:NTUenv}. With metric $g$ fixed, matrices $M_A$ and $M_B$ are optimized to make them the best approximation to the untruncated/exact product $R_A R_B^T$.
For a fixed $M_B$ the cost function \eqref{cost} becomes a quadratic form in $M_A$:
\be 
F_A\left(M_A\right) = M_A^\dag g_A M_A - M_A^\dag J_A - J_A^\dag M_A + F(0),
\ee 
where $g_A$ and $J_A$ depend on the fixed $M_B$, see Figs. \ref{fig:NTUg}(b) and (c). The matrix is optimized as
\be 
M_A={\rm pinv}\left(g_A\right)J_A,
\ee 
where tolerance of the pseudo-inverse can be adjusted to minimize 
$F_A\left[{\rm pinv}\left(g_A\right)J_A\right]$. Thanks to the exactness of $g$ in NTU, the optimal tolerance is usually
in the range $10^{-12},...,10^{-8}$. This optimization of $M_A$ is followed by a similar optimization of $M_B$. The optimizations are repeated in a loop,
\be 
\rightarrow M_A \rightarrow M_B \rightarrow,
\ee 
until convergence of a relative NTU error: $F\left(M_AM_B^T\right)/F(0)$. In general the converged error is non-zero due to the necessary truncation of the bond dimension.

The converged error is used as a criterion to terminate the time evolution. More specifically, we use 
\be 
\delta = dt^{-1} \sqrt{F\left(M_AM_B^T\right)/F(0)}
\label{delta}
\ee
to measure accuracy of the Trotterized evolution. It should be independent of the time step for small enough $dt$. The square root makes $dt\cdot\delta$ a rough estimator of a relative error of the wavefunction inflicted by the Trotter gate and as such also of an error of its expectation values. In the worst case scenario, where the gate errors accumulate in an additive way, a typical relative error after time $t$ is proportional to $t\cdot\delta$, where $\delta$ is averaged over the evolution time.

Except for SVD of small matrices, $R_AR_B^T$ and $M_AM_B^T$, the $\propto D^8$ operations -- required to simulate time evolution in the NTU scheme -- are fully parallelizable. A potential bottleneck is calculation of expectations values that requires the corner transfer matrix renormalization group \cite{Orus_review_14}. However, the expectation values usually do not need to be evaluated after every time step and they may not require the same precision as stable evolution with the full update (FU) scheme.

\bibliography{ref.bib} 

\end{document}